\documentclass[]{aa}
\usepackage{psfig}
\usepackage{subfigure}
\usepackage{txfonts}
\begin{document}

\title{Binary star formation from ring fragmentation}

\author{D.\,A.\,Hubber \and A.\,P.\,Whitworth}

\offprints{David.Hubber@astro.cf.ac.uk}

\institute{School of Physics \& Astronomy, Cardiff University,
5 The Parade, Cardiff, CF24 3YB, UK}

\date{11/03/05}

\abstract{
We present a simple model of binary star formation based on 
the assumption that rotating prestellar cores collapse to 
form rings and these rings then fragment into protostars. 
We assume that each ring spawns a small number (${\cal N} 
\leq 6$) of protostars, and that the
condensation of the protostars is sufficiently rapid
that they can subsequently be treated as point masses.

$\;\;\;$The first part of the model 
is therefore to simulate the dynamical evolution of a ring 
of ${\cal N}$ stars and record the properties of 
the single stars, binaries and higher multiples that form 
as a result of the dissolution of the ring. The masses of 
the individual stars in a ring are drawn from a log-normal 
distribution with dispersion $\sigma_{\log M}$. This part of 
the model is perfomed for many different realizations of 
the ring, to obtain good statistics. It can be formulated 
using dimensionless variables and immediately yields the 
overall multiplicity.

$\;\;\;$The second part of the model is to convolve the results of 
these dimensionless simulations, {\it first} with the distribution 
of core masses, which yields the distributions of multiplicity,
mass ratio and eccentricity, as a function of primary mass; and 
{\it second} with the distribution of core angular momenta, 
which yields the distributions of semi-major axis and period, 
again as a function of primary mass.

$\;\;\;$Using the observed distribution of core masses, and a 
distribution of core angular momenta which is consistent with the 
observations, our model is able to reproduce the observed IMF, the 
observed high multiplicity frequency of pre-Main Sequence stars, 
the observed distribution of separations, and -- for 
long-period systems -- the observed distributions of eccentricity 
and mass-ratio, provided we invoke ${\cal N} = 4\;{\rm or}\;5$ 
and $\sigma_{\log M} = 0.6$.

$\;\;\;$We presume that for short-period 
systems the distributions of eccentricity and mass-ratio are 
modified by the dissipative effects of subsequent tidal 
interaction and competitive accretion; and that the reduced 
multiplicity frequency in the field, compared with young clusters, 
is the result of dynamical interactions between stars formed in 
different cores but the same cluster, following ring dissolution. 
Further numerical experiments are required to explore the 
consequences of such interactions.

}

\maketitle


\section{Introduction} \label{S:INTRO}

One of the main problems in star formation is to explain 
the wide range
of scales over which multiple systems form.  The observed 
distribution of binary separations (e.g Duquennoy \& Mayor 1991,
hereafter DM91; Fischer \& Marcy 1992, hereafter FM92) extends 
from $10^{-2}\,{\rm AU}$ to $10^5\,{\rm AU}$. The explanation 
for this large range of separations must lie in the details of 
the star formation process.  In the field, the multiplicity 
of mature solar-mass stars is around $60\%$ (DM91); and in 
some star formation regions, the multiplicity of 
young solar-mass objects is at least $80\%$, and possibly higher 
(Ghez et al 1993; Reipurth \& Zinnecker 1993; Simon et al 1992;
Patience et al. 2002; Duch$\hat{\rm \bf e}$ne et al 2004).  
This implies that the genesis of multiple systems is an intrinsic 
part of the star formation process.  A prestellar core is presumed 
to collapse and fragment to produce a dense ensemble of protostars. 
Interactions between these protostars and the ambient gas then 
determine their final masses, which ones end up in multiple 
systems, and their orbital parameters.

Understanding how cores fragment and how many objects are
produced requires 3-D hydrodynamical simulations, which
are (a) very computer intensive, and (b) highly dependent
on the input physics and initial conditions of the cores 
(e.g. Tohline 2002). Few simulations can be
performed, and the statistical properties of the resulting
protostars are therefore poorly constrained. Also, such
simulations are not able to follow star formation to completion,
so the final fate of a star-forming core and the stars
it produces is not completely resolved. In particular, the
orbital parameters of the resulting multiple systems are unlikely 
to have reached their final values when a 3-D hydrodynamic 
simulation is terminated.


\begin{figure*} 
\centerline{\psfig{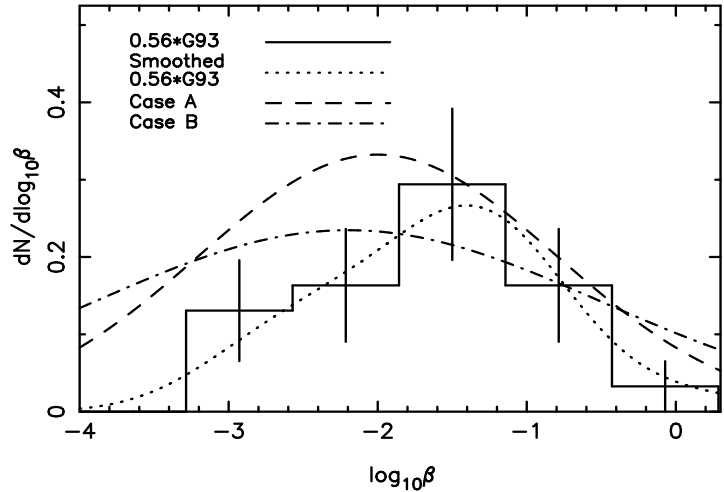}}
\caption{The histogram and the dotted curve represent the distribution 
of observed $\beta$-values from Goodman et al. (1993), scaled to 56\% 
(since they only measured rotation for 24 out of 43 cores). The dashed 
line represents the log-normal distributions of Case A 
($\;\overline{\log \beta} = -2.0$, $\;\sigma_{\log \beta} = 1.2$) and 
the dash-dot line represents the log-normal distribution of Case B 
($\;\overline{\log \beta} = -2.2$, $\;\sigma_{\log \beta} = 1.7$).}
\label{FIG:BETA}
\end{figure*}


\subsection{Previous ${\cal N}$-body work}

A complementary approach to hydrodynamic simulations is to use 
an ${\cal N}$-body code to follow the ballistic evolution of a 
system of protostars, treating them as point masses, and thereby 
ignoring complicated gas-dynamical processes like fragmentation,
 merger and accretion. This approach has been pioneered by Kroupa 
and co-workers (Kroupa 1995a,b; Kroupa \& Bouvier 2003a,b; 
Kroupa et al. 2003), and by Sterzik \& Durisen (1998, 2003), with 
a view to explaining the observed statistics of binary stars.
The work of Sterzik \& Durisen is similar to the present 
study, in that it is mainly concerned with the origin of the 
primordial binary properties in small-${\cal N}$ clusters (or 
subclusters) of protostars. In Kroupa's work the emphasis is 
more on how these primordial binary properties are altered 
by subsequent dynamical interactions with other multiple systems 
and single stars in a wider cluster environment.

Sterzik \& Durisen (1998) investigate the binary statistics 
resulting from the dynamical dissolution of small-${\cal N}$ 
clusters. They pick cluster masses from a power-law core mass 
spectrum, and calculate cluster radii from a scaling law of the 
form $R_{_{\rm CLUTER}} \propto M_{_{\rm CLUSTER}}^\xi$ ($\xi 
= 0,\;1\;{\rm or}\;2$). Each cluster contains ${\cal N}$ stars 
(${\cal N} = 3,\;4\;{\rm or}\;5$) with masses picked from a 
prescribed stellar mass spectrum. Initially the stars are 
positioned randomly in the cluster volume, with zero velocity. 
Their ballistic evolution is then followed for many crossing 
times, using an ${\cal N}$-body code, and the properties of the 
resulting multiple systems are recorded. However, this model is 
unable to reproduce the broad distribution of binary semi-major 
axes observed by DM91. 

In a second paper Sterzik \& Durisen (2003) repeat these experiments, 
but now with clusters which initially are oblate and have some 
rotation about their short axis (specifically, $\beta = 0.1$, where 
$\beta$ is the ratio of rotational to gravitational energy). In addition, 
they relax the assumption of constant ${\cal N}$. Instead stars are 
chosen from a prescribed mass distribution until their total mass 
adds up to the preordained mass of the cluster, and this then determines 
${\cal N}$ for that cluster. They are able to reproduce the dependence 
of multiplicity frequency on primary mass, but the distribution of 
semi-major axes is still much narrower than that observed by DM91.

Delgado-Donate et al. (2003) adopt a somewhat different approach,
designed to capture hydrodynamic effects. They model a uniform-density 
core of isothermal gas using Smoothed Particle Hydrodynamics, and place 
5 sink particles (representing protostellar embryos) at random 
positions within the core. Initially each sink contains only $2\%$ 
of the total mass, but subsequently the sinks grow by competitive 
accretion and interact dynamically with one another, to generate a 
mass function which is a good fit to the observed Initial Mass 
Function. The resulting binary systems also have the right distribution 
of eccentricities. However, the distribution of semi-major axes is 
again much narrower than that observed by DM91, and is more similar
to that observed in open clusters (Patience et al. 2002).

Kroupa (1995a) explores the evolution of a cluster of binary systems, 
and concludes that the properties of G Dwarf binaries in the field 
can be reproduced by dynamical interactions between binaries in a 
young cluster. Specifically he identifies a {\it dominant mode cluster}, 
which starts off with 200 binaries formed by random pairings from the 
field IMF, and with periods drawn randomly from a distribution covering 
1 day to $10^9$ days and having no correlation with primary mass. 
Interactions amongst the binaries (termed {\it stimulated evolution}) 
reduces the binary fraction to $\sim 60\%$, by selectively removing 
low-mass companions (i.e. dynamical biasing), and thereby reproduces 
the distributions of period and mass ratio observed in field G Dwarf 
binaries by DM91. 

In Kroupa (1995b), this model is extended to include the 
{\it eigenevolution} of short-period systems (for example, their tidal 
circularization). He shows that stimulated evolution 
does not change the distribution of orbital eccentricities 
significantly (so the observed distribution must be essentially 
primordial). In addition, stimulated evolution does not generate sufficient 
higher multiples by capture to match the observed numbers of triples 
and quadruples. However, it does produce a distribution of mass ratios 
for G Dwarf primaries which is in good agreement with the observations 
of DM91. Kroupa (1995b) also suggests that long-period binaries with 
large eccentricities are somewhat more likely to be disrupted tidally 
than long-period binaries with low eccentricities.

Kroupa \& Burkert (2001) investigate the ballistic evolution of 
clusters comprising 100 or 1000 primordial binaries, with a narrow 
range of initial separations. They find that even under the most 
favourable conditions, dynamical interactions between binaries 
cannot produce a distribution of separations as broad as that 
observed by DM91.The observed separation distribution must 
therefore be set principally by the properties of primordial 
binaries.

Kroupa, Aarseth \& Hurley (2001) show that the Pleiades may have 
evolved from a cluster like the Orion Nebula Cluster (ONC), 
following loss of residual gas and simultaneous stimulated 
evolution. They point out that the primordial binary population 
in the ONC could have been very similar to the pre-Main Sequence 
binary population that is currently observed in Taurus-Auriga. 
Subsequent dynamical interactions in the dense cluster 
environment then changed it into what we see today.

Kroupa \& Bouvier (2003a) show that the basic model developed in 
Kroupa (1995a) can 
not only explain the high multiplicity fraction for pre-Main 
Sequence stars in Taurus-Auriga, but also the apparent paucity 
of Brown Dwarfs (relative to the ONC). However, the basic model 
is seemingly unable to explain the binary properties of brown 
dwarfs: it produces too many stars with brown dwarf companions, 
and too many wide star-BD and BD-BD systems (Kroupa et al. 2003). 
It is also unable to reproduce the number of brown dwarfs per 
star, unless brown dwarfs have different formation mechanisms 
from stars, and possibly even different formation mechanisms in 
different environments (e.g. embryo ejection, photoevaporation, 
etc.; Kroupa \& Bouvier 2003b).

\subsection{Motivation for a new model}

In simulations of core collapse which include rotation (e.g.
Bonnell \& Bate 1994, Cha 
\& Whitworth 2003, Hennebelle et al. 2004), in particular those where 
instability against collapse is triggered impulsively, the core may 
overshoot centrifugal balance and then bounce to form a dense ring, 
which subsequently fragments into multiple protostars. In this paper, 
we investigate the consequences of assuming that this is the 
dominant mechanism by which a core breaks up into individual protostars. 
Specifically, we use an ${\cal N}$-body code to follow the ballistic
evolution of protostars which are initially distributed on a ring, and 
record their final binary statistics. We scale the mass and radius of 
the ring to match the observed distributions of core mass, core radius, 
and core rotation; and we compare the resulting binary statistics 
with observation.  Our approach is similar to that adopted by 
Sterzik \& Durisen (1998, 2003). The main difference -- and the reason 
that we obtain a broader distribution of separations, comparable with 
what is observed -- is that we consider clusters having a distribution 
of $\beta$-values (i.e. a range of rotation rates), whereas Sterzik \& 
Durisen (1998) considered clusters with no rotation, and Sterzik \& 
Durisen (2003) considered clusters with a universal $\beta\;(=0.1)\,$.

In Section \ref{S:COREOBS}, we review the observations of cores which 
provide the input parameters for our model. In Section \ref{S:STAROBS} 
we review the observations of stars and binary systems which our 
model seeks to explain. In Section \ref{S:OUTLINE} we sketch the 
sequence of stages through which the model is developed. In Section 
\ref{S:DIMSIM}, we present 
the results of dimensionless simulations of ring dissolution.
In Section \ref{S:CONVMASS}, we scale the dimensionless results 
by convolving them with the core mass spectrum and obtain the 
distributions of multiplicity, eccentricity and mass ratio as a 
function of primary mass. In Section \ref{S:CONVBETA}, we scale the 
results, by convolving them with the distribution of core angular 
momenta, and obtain the distribution of semi-major axes, as 
a function of primary mass. In Section \ref{S:CONCLUSIONS}, 
we summarize our main conclusions.


\section{Observations of cores} \label{S:COREOBS}

In this section we review briefly the observations of cores which 
provide the input parameters for our model, namely the distribution 
of core masses (Section \ref{SS:COREMASS}), the mass-radius relation 
for cores (Section \ref{SS:CORERADIUS}), and the distribution of core 
rotation rates (Section \ref{SS:COREBETA}).


\subsection{Core masses} \label{SS:COREMASS}

Motte et al. (2001) have measured the core-mass spectrum in Orion
B and fitted it with a two part power law:
\begin{equation} \label{EQN:CMS}
\frac{d{\cal N}_{_{\rm CORE}}}{dM_{_{\rm CORE}}} =
\left\{ \;\; \begin{array}{ll}
k_{_1} M_{_{\rm CORE}}^{-1.5}, \;\;\;\; & 
 M_{_{\rm MIN}} \leq\ M_{_{\rm CORE}} \leq   M_{_{\rm KNEE}}\,; \\
 \; & \; \\
k_{_2} M_{_{\rm CORE}}^{-2.5}, & 
 M_{_{\rm KNEE}} \leq\ M_{_{\rm CORE}} \leq   M_{_{\rm MAX}} \,.
\end{array} \right .
\end{equation}
where $M_{_{\rm MIN}}=0.5\,{\rm M}_{\odot}$,
$M_{_{\rm KNEE}}=1.0\,{\rm M}_{\odot}$
and $M_{_{\rm MAX}}=10.0\,{\rm M}_{\odot}$.
Similar results have been reported by Johnstone et al. (2001) for 
Orion B, by Motte, Andr\'e \& Neri (1998) and Johnstone et al. (2000) 
for $\rho$ Ophiuchus, and by Testi \& Sargent (1998) for Serpens. We 
shall use Eqn. (\ref{EQN:CMS}) in our model
but we extend the core mass spectrum up to 
$M_{_{\rm MAX}}=20.0\,{\rm M}_{\odot}$.

\subsection{Core radii} \label{SS:CORERADIUS}

We shall assume that the initial core radii are given by the
scaling relations
\begin{equation} \label{EQN:Larson}
R_{_{\rm CORE}}(M_{_{\rm CORE}}) = \left\{ \;\;\begin{array}{ll} 
0.1\,{\rm pc} \, \left( M_{_{\rm CORE}}/{\rm M}_\odot \right), & 
 M_{_{\rm CORE}} < {\rm M}_\odot \,; \\
 \; & \; \\
0.1\,{\rm pc} \, \left( M_{_{\rm CORE}}/{\rm M}_\odot \right)^{1/2}, 
\;\; &  M_{_{\rm CORE}} > {\rm M}_\odot \,;
\end{array}
\right .
\end{equation}
(cf. Larson 1981, Myers 1983). Here, the low-mass regime 
applies to cores whose support is dominated by thermal 
pressure, and the high-mass regime applies to cores whose 
support is primarily from turbulence. Strictly speaking there is a 
range of core radii at any given core mass, but since we are trying 
to keep the number of free parameters in the model to a minimum, we 
neglect this.


\begin{figure*}
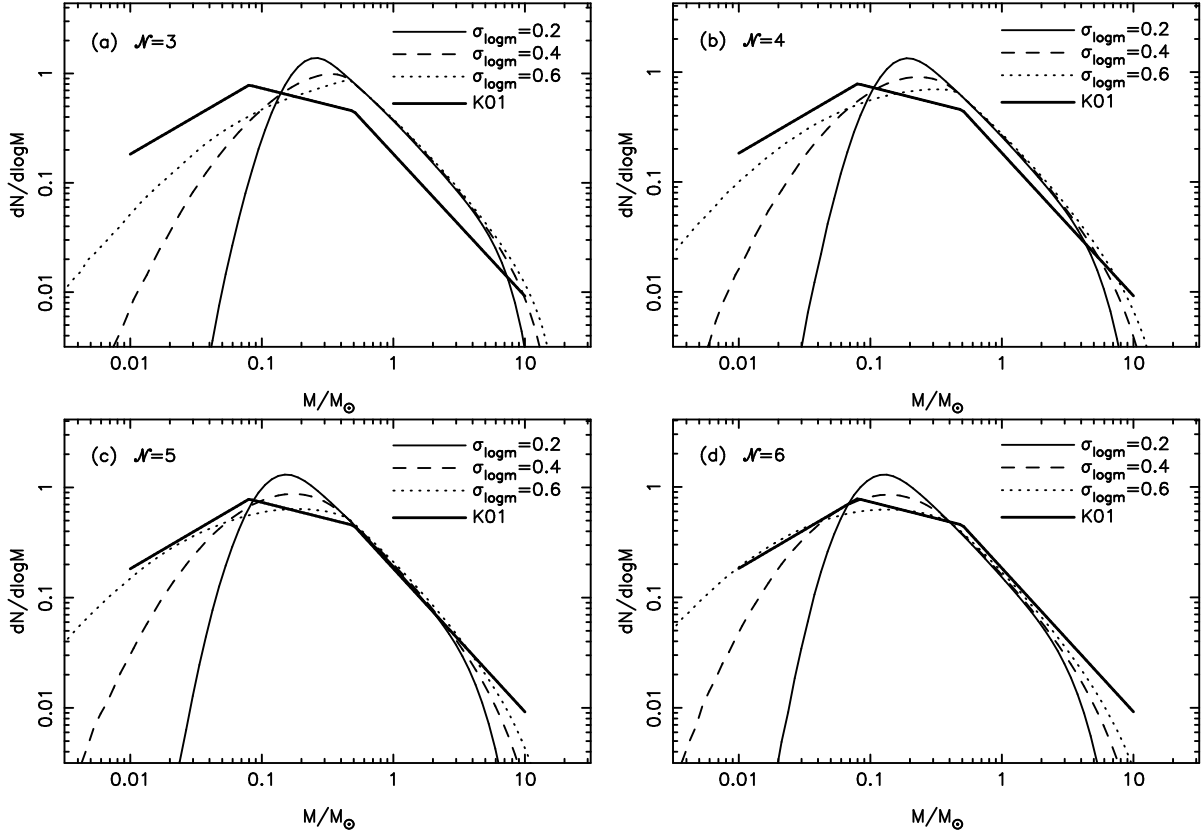
 
\centerline{\psfig{figure=2428IMF1.ps,height=5.5cm,width=7.8cm,angle=270}
\,\,\,\,\psfig{figure=2428IMF2.ps,height=5.5cm,width=7.8cm,angle=270}}
\centerline{\psfig{figure=2428IMF3.ps,height=5.5cm,width=7.8cm,angle=270}
\,\,\,\,\psfig{figure=2428IMF4.ps,height=5.5cm,width=7.8cm,angle=270}}
\caption{(a) The model IMF for ${\cal N} = 3$ with $\sigma_{\log M} = 
0.2\;({\rm thin\;solid\;line}),\;0.4\;({\rm dashed\;line})\;{\rm and}\;0.6\;
({\rm dotted\;line})$. (b) as (a) but for ${\cal N} = 4$. (c) as (a) but 
for ${\cal N} = 5$. (d) as (a) but for ${\cal N} = 6$. The observed IMF 
(Kroupa, 2001) is shown as a heavy solid line on each panel.}
\label{FIG:IMF}
\end{figure*}



\subsection{Core rotation} \label{SS:COREBETA}

Goodman et al. (1993; hereafter G93) have measured the velocity
profiles across a sample of 43 pre-stellar cores. They find 
statistically significant velocity gradients across many of the 
cores, and from these velocity gradients they calculate the ratio 
of rotational to gravitational potential energy, $\beta$, on the 
asumption of uniform global rotation, for 24 of the cores they 
observe. The distribution of these $\beta$-values, normalized 
to $24/43 \simeq 56\%$, is represented by the histogram on 
Fig.~\ref{FIG:BETA}, and also by the dotted line. The dotted line 
is obtained by smoothing the individual $\beta$ values with a 
gaussian kernel having $\sigma_{\log\beta} = 0.5$,
\begin{equation}
\left. \frac{d{\cal N}_{_{\rm CORE}}}{d\log \beta} \right|_{_{\rm OBS}} 
= \sum_i \left\{ \frac{1}{(2\pi)^{1/2}\sigma_{\log\beta}} \exp \left[ 
\frac{-\,(\log\beta_i - \log\beta)^2}{2 \sigma_{\log\beta}^2} \right]\right\}.
\end{equation}
and is intended to mitigate the effects of arbitrarily binning a very small 
number of data points. However, due to low number statistics and the 
fact that only large projected velocity gradients can be measured, the 
full distribution of $\beta$-values is not well constrained by the 
observations of G93.

Burkert \& Bodenheimer (2000) have pointed out that the velocity 
gradients observed by G93 can also be explained by turbulent 
motions in cores, rather than uniform global rotation. By 
modelling the turbulence as a Gaussian random velocity field 
with power spectrum $P(k) \propto k^{-n}$ with $3 \leq n \leq 4$, 
they show that the resulting distribution of $\beta$-values 
fits the G93 observations well, and is approximately log-normal 
(see their Figure 3, lower left panel).

Therefore we assume a log-normal distribution of $\beta$-values:
\begin{equation} \label{EQN:BETAFIT}
\frac{d {\cal N}_{_{\rm CORE}}}{d\log \beta} \;\propto\; \exp 
\left( \frac{-(\log \beta - \overline{\log \beta})^2}
{2 \sigma^2_{\log \beta}} \right) \,,
\end{equation}
Since G93 only determined $\beta$ for 24 of the 43 prestellar cores 
they observed, the observed $\beta$ distribution on Fig.~\ref{FIG:BETA} 
represents only $56\%$ of all the cores, and we therefore have some freedom 
in choosing the parameters $\overline{\log \beta}$ and $\sigma_{\log \beta}$ 
for the full distribution in Eqn. (\ref{EQN:BETAFIT}). We presume that 
some of the cores for which $\beta$ could not be determined were observed 
with inadequate resolution and/or from an unhelpful viewing direction 
(i.e. close to the angular momentum vector). However, this cannot account for 
all the non-determinations, and we assume that the majority of the 
non-determinations have $\beta$ values lower than those that are 
determined. Therefore we must invoke an overall distribution which 
contains the observed distribution, but extends to lower $\beta$ 
values. We consider two possibilities.

In Case A we adopt $\overline{\log \beta} = -2.0$ and 
$\sigma_{\log \beta} = 1.2$ (dashed curve on Fig.~\ref{FIG:BETA}). 
This is the less extreme possibility, in the sense that (a) it is 
easily compatible with the constraint of containing the observed 
distribution, and (b) it has most of the remaining 
$44\%$ of $\beta$ values below, but only just below, the observed ones. 
It is therefore also our prefered possibility. In Section \ref{S:CONVBETA} 
we show that it yields a distribution of separations similar to that of the 
pre-Main Sequence binaries collated by Patience et al. (2002).

In Case B we adopt $\overline{\log \beta} = -2.2$ and 
$\sigma_{\log \beta} = 1.7$ (dot-dash curve on Fig.~\ref{FIG:BETA}). 
This is the more extreme possibility, in the sense that (a) it is 
only just compatible with the constraint of containing the observed 
distribution and (b) it has most of the 
remaining $44\%$ of $\beta$ values not just below, but well below, 
the observed ones. In Section \ref{S:CONVBETA} we show that it yields 
a distribution of separations similar to that of the Main Sequence 
G-dwarf binaries in the field.

We assume that $\beta$ is not correlated with core mass, as indicated 
by G93 (their Fig. 13(b)).


\begin{figure*}
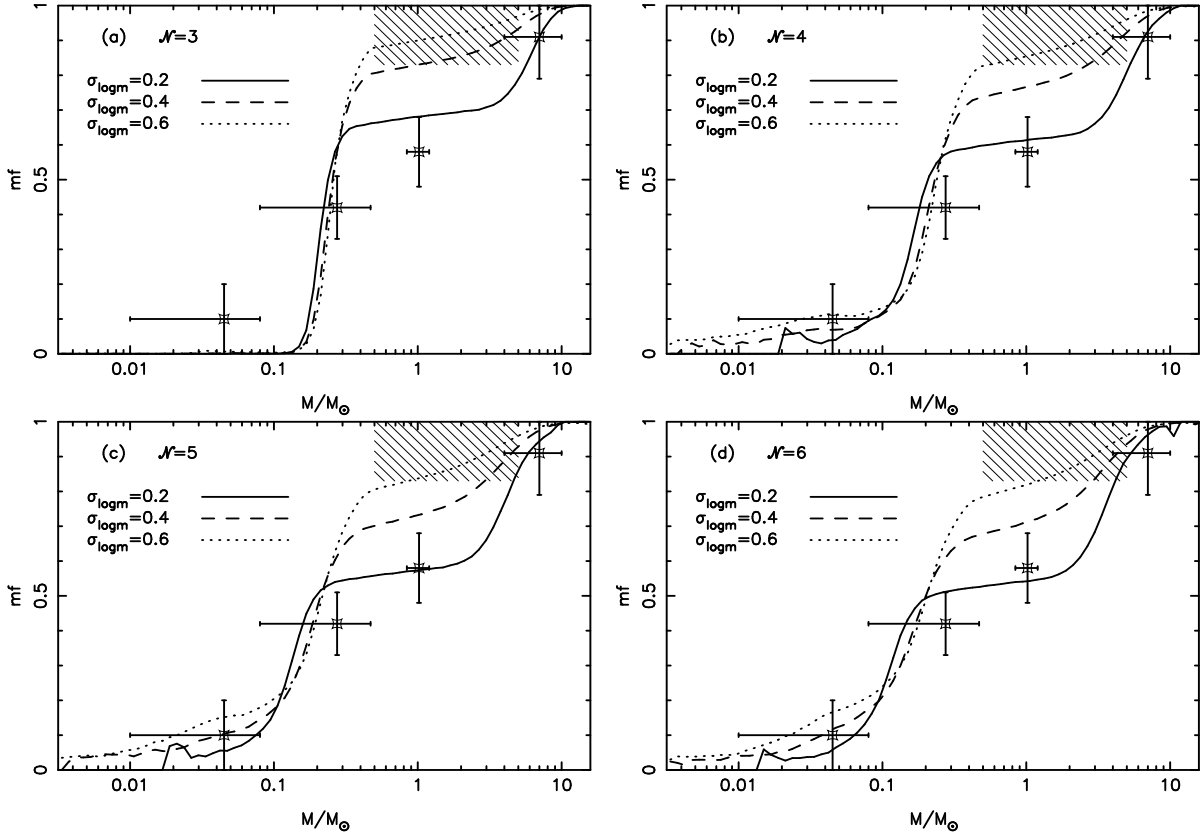
 
\centerline{\psfig{figure=2428MF1.ps,height=5.5cm,width=7.8cm,angle=270} 
\,\,\,\,\psfig{figure=2428MF2.ps,height=5.5cm,width=7.8cm,angle=270}}
\centerline{\psfig{figure=2428MF3.ps,height=5.5cm,width=7.8cm,angle=270}
\,\,\,\,\psfig{figure=2428MF4.ps,height=5.5cm,width=7.8cm,angle=270}}
\caption{The multiplicty frequency as a function of primary mass for (a) 
${\cal N} = 3$, (b) ${\cal N} = 4$, (c) ${\cal N} = 5$, and (d) ${\cal N} = 
6$, all with $\sigma_{\log M} = 0.2,\;0.4\;\&\;0.6$. The four plotted points 
with error bars are observational values taken from Mart\'in et al. (2000), 
FM92, DM91 and Shatsky \& Tokovinin (2002). The hashed box represents 
the extrapolated multiplicty of PMS stars (Patience et al. 2002)}
\label{FIG:MULT}
\end{figure*}



\section{Observations of stars and binary systems} \label{S:STAROBS}

In this section we review briefly the observations of stars and multiple 
systems which our model seeks to explain, namely the stellar Initial 
Mass Function (Section \ref{SS:STARIMF}), stellar multiplicity (Section 
\ref{SS:STARMULT}), the binary separation distribution (\ref{SS:STARPERIOD}), 
the binary eccentricity distribution (\ref{SS:STARe}) and the 
binary mass-ratio distribution (\ref{SS:STARq}). We stress that none 
of these distributions is used as input to the model. They are used solely 
for comparison with the predictions of the model.


\subsection{The Initial Mass Function (IMF)} \label{SS:STARIMF}

We will adopt the IMF determined by Kroupa (2001), i.e.
\begin{equation} \label{EQN:KroupaIMF} 
\frac{d{\cal N}_{_\star}}{dM_{_\star}} \propto 
M_{_\star}^{-\alpha}\,dM_{_\star} \,,
\end{equation}
where
\begin{equation} \begin{array}{ll}
\alpha_0 = +0.3 \pm 0.7, \;\;\;\; & 
 0.01\,{\rm M}_\odot \leq M_{_\star} < 0.08\,{\rm M}_\odot \,; \\
\alpha_1 = +1.3 \pm 0.5, & 
 0.08\,{\rm M}_\odot \leq M_{_\star} < 0.50\,{\rm M}_\odot \,; \\
\alpha_2 = +2.3 \pm 0.3, & 
 0.50\,{\rm M}_\odot \leq M_{_\star} < 1.00\,{\rm M}_\odot \,; \;\;{\rm and} \\
\alpha_3 = +2.3 \pm 0.7, & 
 1.00\,{\rm M}_\odot \leq M_{_\star} \,.
\end{array} \end{equation}
This IMF is plotted on all four panels of Figure \ref{FIG:IMF} 
(heavy solid lines).


\subsection{Stellar multiplicity} \label{SS:STARMULT}

There are many different measures of multiplicity in use (e.g. 
Reipurth \& Zinnecker, 1993; Goodwin et al., 2004b), but we will 
limit our discussion to the multiplicity frequency, ${\bf mf}$. 
If the total number of singles is $S$, the total number of 
binaries is $B$, the total number of triples is $T$, the 
total number of quadruples is $Q$, etc., then
\begin{equation} \label{EQN:MF}
{\bf mf} = \frac{B + T + Q + ...}{S + B + T + Q + ...}
\end{equation}
and gives the fraction of systems which is multiple. ${\bf mf}$ is 
more stable than the other measures, in the sense that it does not 
need to be revised if a multiple system is found to have additional 
components, only if a single has to be reclassified as a multiple. 

Furthermore ${\bf mf}$ can be defined as a function of primary mass,
\begin{equation} \label{EQN:MFM}
{\bf mf}(M_{_1}) = \frac{B(M_{_1}) + T(M_{_1}) + Q(M_{_1}) + ...}
{S(M_{_1}) + B(M_{_1}) + T(M_{_1}) + Q(M_{_1}) + ...} \,,
\end{equation}
where $S(M_{_1})$ is now the number of single stars of mass $M_{_1}$, 
$B(M_{_1})$ is the number of binaries having a primary of mass 
$M_{_1}$, and so on. 

For Main Sequence stars in the field, the observed values of 
${\bf mf}(M_{_1})$ are $0.10 \pm 0.10$ for primaries in 
the mass range $(0.01\,{\rm M}_\odot \leq M_{_1} \leq 0.08\,{\rm M}_\odot)$ 
(Mart\'in et al. 2000), $0.42 \pm 0.09$ for primaries in the mass range
$(0.08\,{\rm M}_\odot \leq M_{_1} \leq 0.47\,{\rm M}_\odot)$ (FM92), $0.58 \pm 0.10$ for primaries in the mass range $(0.84\,
{\rm M}_\odot \leq M_{_1} \leq 1.20\,{\rm M}_\odot)$ (DM91) and $0.91 \pm 0.12$ for primaries in the mass range 
$(4.0\,{\rm M}_\odot \leq M_{_1} \leq 10.0\,{\rm M}_\odot)$ (Shatsky \& 
Tokovinin 2002). These observational points are plotted with error bars 
on Fig. \ref{FIG:MULT}.

For pre-Main Sequence (PMS) stars the situation is less clear, 
because observations of PMS binaries only cover a limited range of 
separations. However, in those separation ranges where PMS binaries 
can be observed, the multiplicity frequency appears to be significantly 
higher than for Main Sequence field stars in the same separation ranges. 
The compilation of Patience et al. (2002) suggests that for PMS 
primaries in the mass range $0.5\,{\rm M}_\odot < M_1 < 5\,{\rm M}_\odot$ 
the multipilicity frequency is in the range $0.83 < {\bf mf} < 1.00$. 
This is shown as a hatched region on Fig.~\ref{FIG:MULT}.


\subsection{The binary period and separation distributions} 
\label{SS:STARPERIOD}

DM91 have measured the binary
properties of a complete sample of multiple systems in the solar 
neighbourhood having Main Sequence G-dwarf primaries. They find 
the distribution of periods to be approximately log-normal, i.e.
\begin{equation}
\frac{d {\cal N}_{_\star}}{d\log P_{\rm d}} \;\propto\; \exp \left\{ -\, 
\frac{\left(\log P_{\rm d} - \overline{\log P_{\rm d}}\right)^2}
{2 \sigma^2_{\log P}} \right\} \,,
\end{equation}
where $P_{\rm d}$ is the period in days and `$\log$' is to the base 10, 
$\overline{\log P_{\rm d}}=4.8$, and $\sigma_{\log P}=2.3$.
The corresponding distribution of separations for multiple systems 
having Main Sequence G-dwarf primaries is then also
log-normal (e.g. Ghez et al. 1993) with $\overline{\log a_{_{\rm AU}}}=1.44$
and $\sigma_{\log a}=1.53$; Figure \ref{FIG:SMA}a illustrates this 
distribution. 

FM92 find a similar
log-normal distribution of periods and separations for multiple systems 
in the solar neighbourhood having Main Sequence M-dwarf primaries.

Pre-Main Sequence binaries (Simon et al 
1992; Ghez et al 1993; Reipurth \& Zinnecker 1993; Patience et al 2002) 
have a distribution of separations that can again be approximated as 
log-normal over a limited range of separations, but it is
different from that for Main-Sequence binaries 
in being shifted to larger separations, and somewhat narrower, 
with $\overline{\log a_{_{\rm AU}}} = 1.8 \pm 0.2$ and $\sigma_{\log a} 
\sim 1.36$ (Patience et al 2002), as illustrated on Figure \ref{FIG:SMA}b.



\subsection{The binary eccentricity distribution} \label{SS:STARe}

DM91 find that for binaries with Main Sequence G-dwarf primaries, 
the eccentricity distribution depends on the period. For long-period 
systems ($P > 10^4\,{\rm days}$, $a > 10\,{\rm AU}$), the eccentricty 
distribution is approximatey thermal (i.e. $d{\cal N}_{_\star}/de = 
2e$, Valtonen \& Mikkola 1991); the DM91 results for long-period 
binaries are shown on Fig.~\ref{FIG:eANDq2}a. 
For short-period binaries with Main Sequence G-dwarf primaries 
($P < 10^4\,{\rm days}$), the eccentricities tend to be significantly 
lower, with a distribution peaked around $e \simeq 0.2$. In addition, 
there appears to be an upper limit on the eccentricity, 
$e_{_{\rm MAX}}(P)$, which decreases with decreasing $P$ and approaches 
zero for $P \la 10\,{\rm days}$.

The data available for binaries with Main Sequence primaries of other 
spectral types (i.e. M-dwarfs and K-dwarfs) are limited, particularly 
for systems with long periods, but the overall distribution of 
eccentricity with period appears to be broadly similar to that for 
binaries with G-dwarf Main Sequence  primaries. In particular, the 
upper limit on the eccentricity, $e_{_{\rm MAX}}(P)$, decreasing with 
decreasing $P$ and approaching zero for $P \la 10\,{\rm days}$, appears 
to apply to all Main Sequence binaries. This upper limit on $e$ is 
normally attributed to tidal circularization of close orbits.

For Pre-Main Sequence binaries the data on eccentricity is limited 
to short-period systems ($P < 10^4\,{\rm days}$). Again there appears 
to be an upper limit on the eccentricity, $e_{_{\rm MAX}}(P)$, which 
decreases with decreasing $P$ (Mathieu 1994). However, this limit is 
somewhat larger than that for binaries with G-dwarf Main Sequence primaries, 
and it only approaches zero for $P \la 3\,{\rm days}$. Again, this is 
consistent with a picture in which close systems are circularized 
tidally; in pre-Main Sequence systems there has been less time 
for the process to work.


\subsection{The binary mass-ratio distribution} \label{SS:STARq}

The binary mass ratio, $q$, is defined by $q = M_{_2} / M_{_1}\,$, 
where $M_{_1}$ is the primary mass, so necessarily $q\leq 1$.

DM91 find that for binaries having Main Sequence G-dwarf primaries, 
the distribution of $q$-values is dependent on the period. For 
long-period systems ($P > 10^4\,{\rm days}$), the $q$-distribution 
has a significant peak at around $q = 0.3$ (see Figure 
\ref{FIG:eANDq2}b). For short-period systems, a much flatter 
distribution is observed, gently rising towards $q = 1$ (Mazeh et 
al. 1992). 

Mass ratios have also been determined for binaries having 
Main Sequence M-dwarf primaries by FM92. However, there are too 
few systems to reveal any clear dependence on period, and the 
sample is incomplete for $q < 0.4$. For the whole sample, 
the distribution of mass ratios is consistent with being flat 
in the range $q \ga 0.5$, but there is the suggestion of a 
decrease for lower $q$-values. We note that for M-dwarf 
primaries these low $q$-values ($q \la 0.5$) usually correspond 
to brown dwarf companions.

Woitas et al. (2001) have determined mass ratios for pre-Main 
Sequence binaries with primary masses corresponding roughly to 
G- and M-dwarf Main Sequence stars. The mass ratios depend 
somewhat on whose evolutionary tracks are used to determine 
individual masses, but again the results are consistent with 
a flat distribution at large mass ratios ($q \ga 0.5$). There 
appears to be a decrease for $q \la 0.3$, but this is probably due 
to incompleteness.



\begin{figure*}
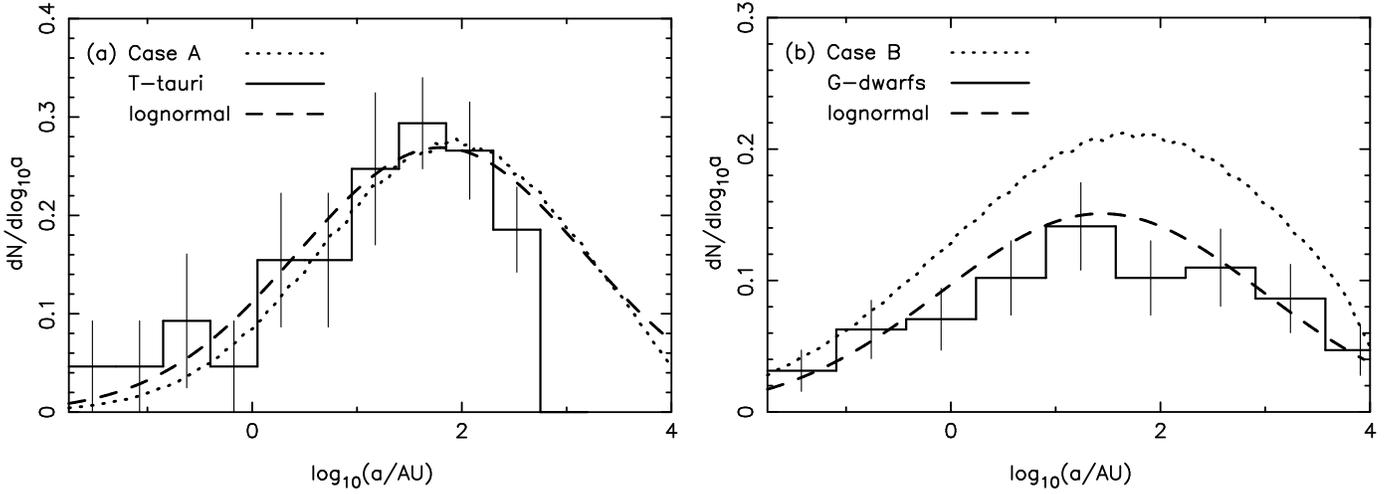
 
\centerline{\psfig{figure=2428Psma.ps,height=6.5cm,width=9.0cm,angle=270}
\,\,\,\,\psfig{figure=2428Gsma.ps,height=6.5cm,width=9.0cm,angle=270}}
\caption{(a) The histogram shows the observed distribution of semi-major 
axes for pre-Main Sequence binaries in Taurus, as collated by Patience 
et al. (2002), and the dashed line is their best fit to this 
distribution. The dotted line shows the predictions of our model for 
Case A ($\overline{\log\beta} = -2.0$ and $\sigma_{\log\beta} = 1.2$)
with parameters ${\cal N}=5$ and $\sigma_{\log M}=0.6$. 
The peak and width of the model predictions are fairly close to the 
observations, with a similarly high overall multiplicity. (b) The 
histogram shows the observed distribution of semi-major axes for 
binaries with G-dwarf primaries in the field from DM91, and the dashed 
line is their log-normal fit to this distribution. The dotted line shows 
the predictions of our model for Case B ($\overline{\log\beta} = -2.2$ 
and $\sigma_{\log\beta} = 1.7$) with parameters ${\cal N}=5$ and 
$\sigma_{\log M}=0.6$. The peak and width of the model 
predictions are similar to the observations, but the overall multiplicty 
is higher.}
\label{FIG:SMA}
\end{figure*}



\section{Outline of model} \label{S:OUTLINE}

\subsection{Assumptions and aims} \label{SS:AIMS}

Our model of binary star formation is based on the assumption 
that all cores are rotating, that they collapse and fragment 
via ring formation, and that the resulting protostars then 
interact ballistically to form multiple systems. The aim of 
the paper is to investigate whether this simple model can 
explain the observed multiplicity of stars and their distributions 
of period, separation, eccentricity and mass-ratio.


\subsection{The model} \label{SS:MODEL}

Consider a rotating prestellar core of mass $M_{_{\rm CORE}}$ which 
initially has radius $R_{_{\rm CORE}}$ and ratio of rotational to 
gravitational energy 
\begin{equation}\label{beta2}
\beta \;=\; \frac{R_{_{\rm CORE}}^{3}\,\Omega_{_{\rm CORE}}^{2}}
{3\,G\,M_{_{\rm CORE}}} \;=\; \frac{25\,H_{_0}^2}
{12\,G\,M_{_{\rm CORE}}^3\,R_{_{\rm CORE}}} \ll 1 \,.
\end {equation}
If the core collapses conserving its angular momentum, $H_{_0}$, and then 
bounces to form a centrifugally supported ring, the ring has radius
\begin{equation} \label{EQN:RRING}
R_{_{\rm RING}} \;\simeq\; \beta \, R_{_{\rm CORE}} \,.
\end{equation}

Suppose further that the ring is formed with 
approximately uniform line-density, but then fragments into ${\cal N}$ 
protostars (where we expect ${\cal N}$ to be small). Assume (i) that the 
${\cal N}$ protostars formed from a single core have masses $M_n$ ($n = 
1, 2, ..., {\cal N}$) drawn from a log-normal distribution with standard 
deviation $\sigma_{\log M}$, and normalized so that 
\begin{equation}
\sum_{n=1}^{n={\cal N}}\,\left\{ M_n \right\} \;=\; f\,M_{_{\rm CORE}} \,,
\end{equation}
where $f$ is the fraction of the core mass which is converted into stars
(for simplicity we set $f=1$ here);
(ii) that the protostars are initially distributed round the ring so 
that each protostar occupies a fraction of the circumference proportional
to the protostar's mass; and (iii) that the protostars condense out 
sufficiently fast that we can follow their subsequent dynamics using 
pure ${\cal N}$-body methods.

For fixed ${\cal N}$ and $\sigma_{\log M}$, we first formulate the 
dynamical evolution in dimensionless form and simulate a large number 
of cases to obtain statistically robust distributions of (a) multiplicity, 
(b) orbital eccentricity, $e$, (c) component mass-ratio, $q \equiv M_2/M_1$,
and (d) ratio of orbital separation to ring radius, $a/R_{_{\rm RING}}$.

Then we convolve, {\it first} with the distribution of core masses, 
to obtain the overall stellar initial mass function (IMF) and 
the distributions of multiplicity, eccentricity and mass-ratio 
as a function of primary mass; and {\it second} with the distribution 
of core $\beta$-values to obtain the distribution of separations 
as a function of primary mass $M_1$.

The core mass spectrum is fairly tightly constrained by observation 
(Motte et al. 1998, Testi \& Sargent 1998, Johnstone et al. 2000, 
Motte et al. 2001, Johnstone et al. 2001), and therefore we do not 
adjust it (Section \ref{SS:COREMASS}). Likewise the relation between 
core mass and core initial radius is constrained by 
observations, and we do not adjust it (Section \ref{SS:CORERADIUS}). 
However, the distribution of $\beta$-values is less well constrained, 
and we consider two possible distributions (Section \ref{SS:COREBETA}). 
${\cal N}$ and $\sigma_{\log M}$ are treated as free parameters. 


\begin{figure*}
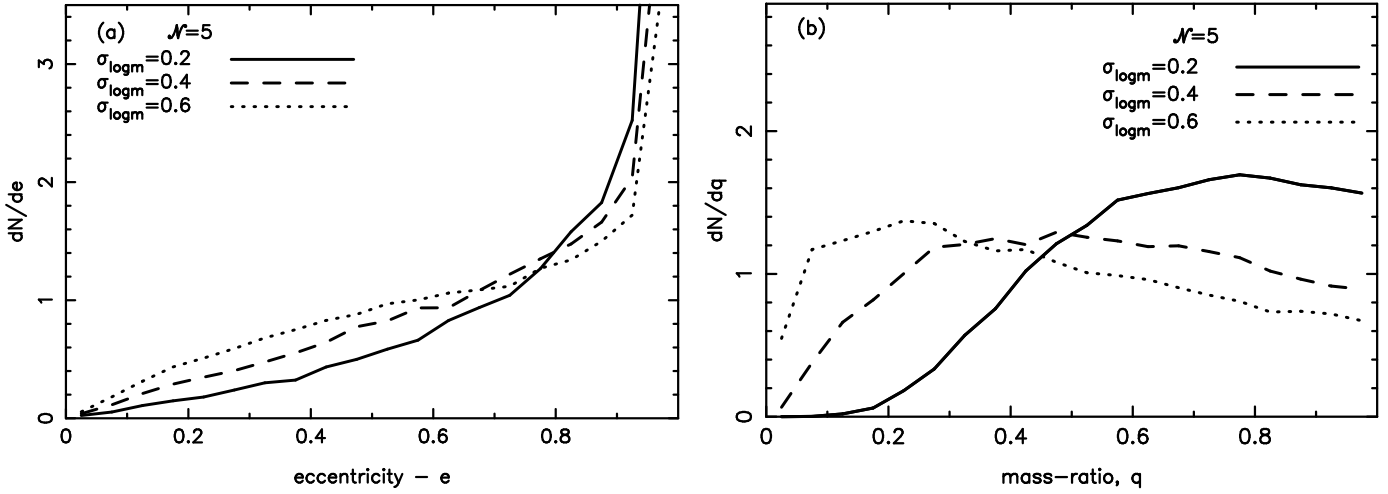
 
\centerline{\psfig{figure=2428dlec.ps,height=6.5cm,width=9.0cm,angle=270}
\,\,\,\,\psfig{figure=2428dlq.ps,height=6.5cm,width=9.0cm,angle=270}}
\caption{ The distributions of (a) eccentricity and (b) mass-ratio
resulting from the dimensionless simulations.
The parameters used are ${\cal N} = 5$ and 
$\sigma_{\log M} = 0.2\;({\rm solid\;line}),\;0.4\;({\rm dashed\;line})
\;\&\;0.6\;({\rm dotted\;line})$.}
\label{FIG:eANDq}
\end{figure*}



\section{Dimensionless simulations} \label{S:DIMSIM}

The first stage in constructing the model is to perform dimensionless 
simulations of low-${\cal N}$ star clusters. The results of these 
simulations can later be scaled up to any mass and size depending on 
the given parameters of the core, i.e. $M_{_{\rm CORE}}$ and $\beta$.


\subsection{Initial Conditions} \label{SS:DIMICS}

We assume that a collapsing core forms a centifugally supported ring 
having radius $R_{_{\rm RING}}$ given by Eqn. (\ref{EQN:RRING}). 
However, for the dimensionless simulations we set $R_{_{\rm RING}} 
= 1$. There are then two free parameters which must be explored in the 
dimensionless simulations. First, we must specify the number of stars in 
the ring, ${\cal N}$. Second, as posited in Section \ref{SS:MODEL}, 
the stellar masses must be drawn at random from a log-normal distribution 
having standard deviation $\sigma_{\log M}$,
\begin{equation} \label{MS}
\frac{d{\cal N}_{_\star}}{d \log M} =\frac{{\cal N}}{(2\pi)^{\frac{1}{2}}
\sigma_{\log M}} \exp \left( \frac{- \left(\log M \right)^2}
{2{\sigma^2_{\log M}}} \right) \,,
\end{equation}
and therefore we must specify $\sigma_{\log M}$. Strictly the mass 
distribution extends from $-\infty$ to $+\infty$ but for computational 
convenience we choose to curtail it at $\pm\,3\,\sigma_{\log M}$. Once 
the ${\cal N}$ stellar masses $M_n$ ($n = 1 \;{\rm to}\; {\cal N}$) 
have been drawn randomally from this distribution, they are re-scaled 
by a factor $g$, $\;M_n \rightarrow \mu_n=gM_n$, so that the total 
system mass is equal to unity, i.e.
\begin{equation} \label{EQN:NORM}
\sum_{n=1}^{n={\cal N}}\,\left\{ \mu_n \right\} \;=\; 1 \,.
\end{equation}
(We note that scaling the stellar masses in this way skews the 
overall distribution of masses slightly; it is no longer precisely 
log-normal. This is not a critical element of the model. The 
explanation is given in the Appendix.)

Next, we must specify the initial positions and velocities of
the stars on the ring.  If a ring having uniform line-density
fragments into ${\cal N}$ stars, then each star $n$ forms from
material in an `angular segment', $\Delta{\theta_n}$, proportional 
to it's mass, i.e.
\begin{equation}
\Delta{\theta_n} = 2\,\pi\,\mu_n \,.
\end{equation}
Thus in circular polar co-ordinates, we put $\theta_1=0$ and
\begin{equation} \label{theta}
\begin{array}{ll}
\theta_n = \theta_{n-1} + \frac{1}{2} \left( \Delta{\theta_{n-1}}
+ \Delta{\theta_n} \right) , &
n = 2,3,.,{\cal N}.
\end{array}
\end{equation}

In order to ensure conservation of linear momentum, each star $n$
must be placed at the centre of mass of the material from 
which it forms, i.e. at radius
\begin{equation}
r_n = \frac{{\rm sin}(\mu_n\,\pi)}{\mu_n\,\pi} \,;
\end{equation}
and then it must be given a circular speed
\begin{equation}
v_n = V\,\frac{{\rm sin}(\mu_n\,\pi)}{\mu_n\,\pi} \,,
\end{equation}
where $V$ must be chosen so that the system is virialized. 
Only part of the angular momentum of the initial ring goes into 
the orbital motion of the stars in the cluster. The remaining 
angular momentum must go into spin of the individual stars and 
their attendant discs, which are not modelled here.


\begin{figure*}
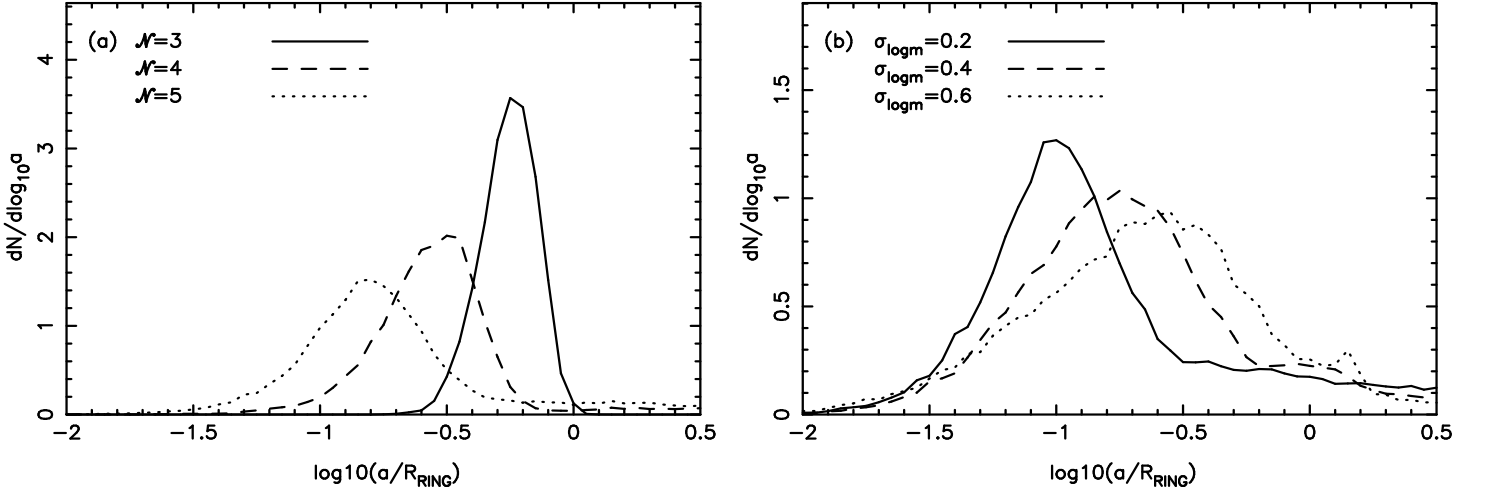

\centerline{\psfig{figure=2428sma1.ps,height=6.5cm,width=9.5cm,angle=270}
\,\,\,\,\psfig{figure=2428sma2.ps,height=6.5cm,width=9.5cm,angle=270}}
\caption{The dimensionless separation distribution for clusters with
a) ${\cal N} = 3,\;4\;\&\;5$ and $\sigma_{\log M} = 0.2$,
and b) ${\cal N} = 6$ and $\sigma_{\log M} = 0.2,\;0.4\;\&\;0.6$.}
\label{FIG:DIMPER}
\end{figure*}


\subsection{Numerical method} \label{SS:DIMNUM}

The ring is then evolved balistically using an adapted 
version of NBODY3, supplied by Sverre Aarseth (e.g. Aarseth 
1999). NBODY3 uses a fourth-order integrator, plus special 
regularisation routines, which are required for close 
encounters.  2-body regularisation is used to treat binaries
and close encounters by transforming the co-ordinates of the
binary components and calculating the motion separately from
the rest of the system (e.g. Aarseth 2001).This technique 
reduces the error generated during close encounters and 
eliminates the need for extra terms such as gravity softening. 
Close interactions between three bodies or more can be treated 
using 3-body, 4-body and chain regularisation. Wider binaries 
and higher order multiple systems such as heirachical triples 
and quadruples are identified by calculating the 2-body 
energies of all the pairs and selecting the most bound pair.

\subsection{Parameters} \label{SS:DIMPARAM}

Simulations are performed for all possible combinations of 
${\cal N}=3,\;4,\;5\;\&\;6$, and $\sigma_{\log M}=0.2,\;0.4
\;\&\;0.6$, and the distributions of multiplicity, semi-major 
axis (as a fraction of the radius of the ring), eccentricity, 
and mass-ratio are recorded.

For each set of parameters (i.e. each pair of ${\cal N}$ and 
$\sigma_{\log M}$ values), a large number of runs is 
required to obtain statistically significant distributions. 
Each set of runs treats $10^5$ stars in total (e.g. for 
${\cal N}$=4, $2.5{\rm x}10^4$ runs are performed). 
Small-N systems usually dissolve in 
a few tens of crossing times (e.g. Van Albada 1968, Sterzik 
\& Durisen 1998), so, to ensure that the majority of 
systems have dissolved at the end of a simulation, each 
realisation is run for about 1000 crossing times. We choose 
a conservative tolerance parameter for the time step, to ensure 
accurate integration. Specifically we require that energy is 
conserved to 1 part in $10^5$ over the entire integration. 



\begin{table} \label{multtable}
\centering
\caption{\small{The numbers of different multiple 
systems produced in the dimensionless simulations}}
\begin{tabular}{|c|c|l|l|l|l|} \hline
& & ${\cal N}=3$ & ${\cal N}=4$ &${\cal N}=5$ &${\cal N}=6$ \\ \hline
$\sigma_{\log m}=0.2$ & $S$ & 33253 & 39711 & 43873 & 46567 \\
                      & $B$ & 33241 & 22506 & 20207 & 19306 \\
                      & $T$ & 88    & 5091  & 5014  & 4394 \\
                      & $Q$ & 0     & 1     & 168   & 409 \\ \cline{2-6}
                 & {\bf mf} & 0.5001 & 0.4100 & 0.3666 & 0.3411 \\ \hline
$\sigma_{\log m}=0.4$ & $S$ & 33231 & 41883 & 45534 & 48053 \\
                      & $B$ & 33228 & 22073 & 18033 & 16670 \\
                      & $T$ & 104   & 4657  & 5764  & 5389 \\
                      & $Q$ & 0     & 0     & 277   & 609  \\ \cline{2-6}
                 & {\bf mf} & 0.5008 & 0.3896 & 0.3459 & 0.3205 \\ \hline
$\sigma_{\log m}=0.6$ & $S$ & 33161 & 40950 & 41598 & 47707 \\
                      & $B$ & 33155 & 23098 & 16955 & 16378 \\
                      & $T$ & 176   & 4282  & 5645  & 5726  \\
                      & $Q$ & 0     & 1     & 333   & 589   \\ \cline{2-6}
                 & {\bf mf} & 0.5013 & 0.4007 & 0.3554 & 0.3223 \\ \hline
\end{tabular}
\end{table}


\begin{figure*}
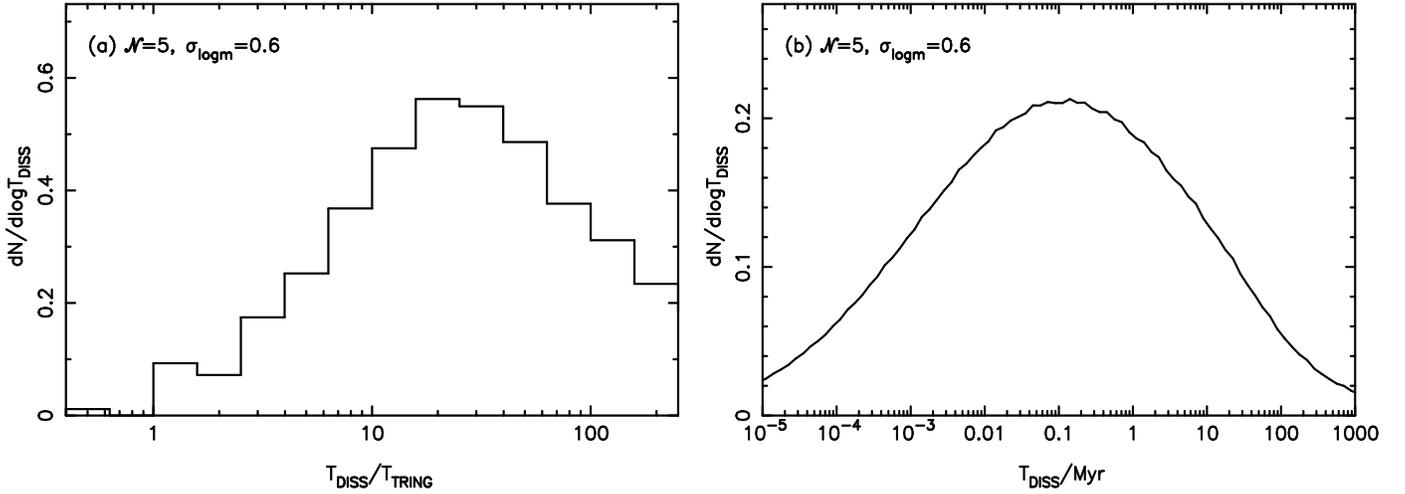
 
\centerline{\psfig{figure=2428dis1.ps,height=6.5cm,width=9.0cm,angle=270}
\,\,\,\,\psfig{figure=2428dis2.ps,height=6.5cm,width=9.0cm,angle=270}}
\caption{a) Dimensionless lifetimes of small-${\cal N}$ rings for 
${\cal N}=5$ and $\sigma_{\log M}=0.6$ b) Lifetimes of all clusters 
with ${\cal N}=5$ and $\sigma_{\log M}=0.6$ when convolved
over the core mass spectrum and the $\beta$ distribution in Myrs.}
\label{FIG:DECAYTIME}
\end{figure*}

\subsection{Dissolution timescale} \label{SS:DIMDISS}

Typically a cluster dissolves after a few tens of dynamical 
times (e.g. Van Albada 1968) leaving ejected singles and a variety 
of multiple systems, i.e. binaries, triples and quadruples.  In 
the present context, we define the dynamical timescale of the initial 
ring to be $T_{_{\rm RING}} = 2\pi \left( R_{_{\rm RING}}^3 / G 
M_{_{\rm CORE}}\right)^{1/2}$; this is roughly equivalent to the 
crossing time for a cluster with isotropic velocity dispersion. 
Figure \ref{FIG:DECAYTIME}a shows the dissolution times, 
$T_{_{\rm DISS}}$, as a function of $T_{_{\rm RING}}$, for the 
case ${\cal N} = 5$ and $\sigma_{{\rm log}M} = 0.6$. About half of 
the rings have dissolved after 20 $T_{_{\rm RING}}$, in agreement 
with previous numerical work.

\subsection{Multiplicity} \label{SS:DIMMULT}

Table \ref{multtable} shows the total numbers of singles, binaries,
triples and quadruples produced and the multiplicity frequencies 
for different values of ${\cal N}$ and $\sigma_{\log M}$. As 
${\cal N}$ is increased, {\bf mf} decreases. The reason for this 
is that a small cluster relaxes by dynamically ejecting stars, 
leaving a stable multiple system. For higher ${\cal N}$, more stars 
need to be ejected as singles before the cluster stabilizes, and this 
increases $S$, thereby reducing {\bf mf}. We do not consider systems 
with very large ${\cal N}$, since such systems are found to produce 
too many singles to match the observations. 
Also hydrodynamical simulations suggest 
that ring fragmentation produces only a small numbers of fragments 
(e.g. Cha \& Whitworth 2003). The effect of changing $\sigma_{\log M}$ 
is quite small compared with changing ${\cal N}$, and is not monotonic.


\subsection{The dimensionless separation distribution} \label{SS:DIMPER}

In Figures \ref{FIG:DIMPER} a \& b, we see that the separation distribution 
has an approximately log-normal form, but for high-${\cal N}$ there 
is an asymmetric tail stretching to large-separations. 

If we fix $\sigma_{\log M}$ and increase ${\cal N}$ (Fig. 
\ref{FIG:DIMPER}a), the peak separation shifts to smaller 
values and the overall distribution becomes broader. 
Increasing ${\cal N}$ tends to 
increase the number of 3-body interactions which must occur before 
a stable multiple is formed, and therefore it tends to increase the 
binding energy of the surviving multiple, i.e. to decrease 
its semi-major axis. 
The greater number of 3-body interactions also produces a wider 
logarithmic range of final separations because it is a stochastic process 
(i.e. one binary system may be stabilized by just two mild 
interactions and another by four violent interactions). 

If we now fix ${\cal N}$ and increase $\sigma_{\log M}$
(Fig. \ref{FIG:DIMPER}b), the separation distibution shifts to larger
separations and becomes somewhat broader.  This is because the number
of 3-body interactions which occur before a stable multiple is
formed is fixed by ${\cal N}$ and is therefore the same.
However, for large $\sigma_{\log M}$, the two most massive stars
which form the final binary contain almost all the mass, and
the remaining stars are so lightweight, that their ejection does
not harden the binary much. Conversely, for small $\sigma_{\log M}$
the two stars which form the final binary are of comparable mass
to those which get ejected, and so their ejection hardens the
remaining binary considerably.

\subsection{Eccentricities} \label{SS:DIMe}

The distribution of eccentricities depends strongly on the 
value of $\sigma_{\log M}$, and much less on the value of ${\cal N}$.
For low $\sigma_{\log M}$, there are few low-eccentricity binaries, 
and the numbers increase monotonically with increasing eccentricity, 
rising rapidly for very high eccentricities ($e > 0.9$). The 
rise at high eccentricities is characteristic of the dissolution of 
2D planar systems; in contrast, the dissolution of 3D systems results 
in a thermal distribution of eccentricities ($d{\cal N}/de = 2e$, e.g. 
Valtonen \& Mikkola 1991).  As we increase $\sigma_{\log M}$, the 
number of high eccentricity binaries decreases and the distribution 
becomes flatter (but not flat, see Fig.~\ref{FIG:eANDq}a).

\subsection{Mass-ratios} \label{SS:DIMq}

The distribution of mass-ratio, $q = M_2/M_1$, is found to be strongly 
dependent on $\sigma_{\log M}$, and only lightly dependent on ${\cal N}$. 
$\sigma_{\log M}$ controls the range of masses possible in a single core 
and thus the possible masses of components in a binary.  If there is 
a low range of masses available, $q$ cannot differ greatly from unity. 
Figure \ref{FIG:eANDq}b shows that for low $\sigma_{\log M}=0.2$, 
most of the binaries have mass-ratios greater than 0.5 with a peak at 
around $q=0.8$.  As $\sigma_{\log M}$ is increased, there are more 
binaries with lower mass-ratios and the peak moves to a lower value 
of $q$.  


\section{Convolving with the core-mass spectrum} \label{S:CONVMASS}

We can now convolve the dimensionless simulations with the core mass 
spectrum to produce an overall distribution of stellar masses (i.e. an 
IMF), the multiplicity as a function of primary mass $M_{_1}$, and the 
distributions of eccentricity $e$ and mass-ratio $q$, as functions 
of $M_{_1}$.


\subsection{The resultant IMF} \label{SS:CMIMF}

First, we look at the IMFs generated by our model, and compare 
them with the observed IMF. The model IMFs, $d{\cal N}_{_\star} 
/ dM_{_\star}$, are given by folding together the core mass function,
$d{\cal N}_{CORE}/dM_{CORE}$ (Equation \ref{EQN:CMS}), 
and the dimensionless stellar mass spectrum, 
$d{\cal N}_{\mu}/d\mu$, as defined in section \ref{SS:DIMICS},
\begin{equation} \label{convolution}
\frac{d{\cal N}_{_\star}}{dM_{_\star}} \,=\, 
\int_{M_{_{\rm CORE,MIN}}}^{M_{_{\rm CORE,MAX}}} \,
{\frac{d{\cal N_{\mu}}}{d\mu} 
\left(\mu=\frac{M_{_\star}}{M_{_{\rm CORE}}} \right) \,
\frac{d{\cal N}_{_{\rm CORE}}}{dM_{_{\rm CORE}}} \,
\frac{dM_{_{\rm CORE}}}{M_{_{\rm CORE}}}} \,.
\end{equation}
Figure \ref{FIG:IMF} shows the IMFs for all combination of 
${\cal N} = 3,\,4,\,5\;\&\;6$ and $\sigma_{\log M} = 0.2,\,0.4 
\;\&\;0.6$. The shape of the IMF is highly dependent on 
$\sigma_{\log M}$. If we were to choose equal-mass stars (i.e. 
effectively $\sigma_{\log M} = 0.0$), the IMF would exactly mimic 
the shape of the core mass spectrum. As we increase $\sigma_{\log M}$, 
the IMF becomes broader. The lowest mass core has mass 
$0.5\,{\rm M}_{\odot}$, so a large value of $\sigma_{\log M}$ 
is required to produce the large observed numbers of low-mass 
stars and brown-dwarves. 
${\cal N}$ has little effect on the shape of the IMF, but 
affects the position of the peak. If we keep $\sigma_{\log M}$ 
constant and increase ${\cal N}$, the overall shape of the IMF 
is roughly constant, but the peak moves to smaller mass. We 
can fit the Kroupa IMF well with ${\cal N} = 5$ and 
$\sigma_{\log M} \sim 0.6$; the ratio of brown dwarfs to stars 
is then $\sim 0.5$.


\subsection{Multiplicity} \label{SS:CMMULT}

Figure \ref{FIG:MULT} shows the multiplicity frequency as a 
function of stellar mass, ${\bf mf}(M_{_1})$, for all combination 
of ${\cal N} = 3,\,4,\,5\;\&\;6$ and $\sigma_{\log M} = 0.2,\,0.4 
\;\&\;0.6$. Overall, the model results have a similar trend to the 
observations, with ${\bf mf}$ increasing from near zero at the 
lowest masses to near unity at the highest masses.

As $\sigma_{\log M}$ is increased, ${\bf mf}$ increases at all 
masses above the peak in the initial mass function. This is 
because, as $\sigma_{\log M}$ is increased, these masses are 
increasingly likely to be the most massive star in the ring, 
and hence increasingly likely to form part of a multiple system 
due to dynamical biasing (McDonald \& Clarke, 1993).

As ${\cal N}$ is increased, ${\bf mf}$ decreases at all masses. 
This is because a small cluster evolves to a stable state (i.e. 
a binary), or a quasi-stable configuration (i.e. a hierarchical 
multiple), by ejecting stars. For higher ${\cal N}$, it is 
necessary to eject more stars before stability, or quasi-stability, 
is reached, and this decreases the overall multiplicity. If there 
is a range of masses, the lower mass stars are ejected preferentially. 

Figure \ref{FIG:MULT} shows that for intermediate stellar masses
($0.4\,{\rm M}_\odot \stackrel{<} {\sim} M_{_*} \stackrel{<}{\sim} 
4\,{\rm M}_\odot$) {\bf mf} is almost independent of primary mass 
$M_1$. This is a direct consequence of using a simple power-law 
core mass function and convolving with a dimensionless distribution. 
For stars of given mass $M_{_\star}$ in the range $0.4\,{\rm M}_\odot$ 
to $4\,{\rm M}_\odot$, there is an approximately constant ratio 
between the number of stars which have formed in a relatively 
low-mass core (and are therefore probably the most massive stars 
in that core and likely to end up as the primary in a multiple 
system) and the number of stars which have formed in a relatively 
high-mass core (and are therefore probably one of the less massive 
stars in that core and unlikely to end up as a primary). This 
approximately constant ratio translates into an approximately 
constant ${\bf mf}$.

For low masses ($M \la 0.2\,{\rm M}_\odot$),
{\bf mf} decreases rapidly with decreasing $M$. Low mass stars
and brown dwarfs are therefore seldom found as primaries in binary
systems, in agreement with observations.  This decrease of 
${\bf mf}$ below ${\sim} 0.2\,{\rm M}_\odot$ would be less severe
if the core mass spectrum were not cut off abruptly below 
$M_{MIN}=0.5 {\rm M}_\odot$ (see Eqn. \ref{EQN:CMS}).

If we consider the multiplicity of pre-Main Sequence stars,
a high multiplicity ($> 0.8$) in the mass range
$0.5\,{\rm M}_\odot \leq M_{_1} < 5.0\,{\rm M}_\odot$ can only 
be realised in our model if $\sigma_{\log M}$ is large ($\sim 0.6$). 
To obtain a good fit to the IMF, we require ${\cal N}=5$ and 
$\sigma_{\log M}=0.6$ (see Section \ref{SS:CMIMF}). 
Therefore we can obtain a good fit to {\it both} the observed IMF, 
{\it and} the observed high ${\bf mf}$ for PMS stars, with 
${\cal N}=5$ and $\sigma_{\log M}=0.6$.

However, there exists no parameter set that results in a 
satisfactory fit to the observed IMF, and reproduces the 
${\bf mf}$ of mature field stars.  For example, to obtain a 
${\bf mf}$ of about 0.6 for G-dwarf primaries requires a low 
$\sigma_{\log M} \simeq 0.2$, but this produces a poor IMF 
which has too few brown dwarfs. This inconsistency can be 
resolved, if we retain $\sigma_{\log M}=0.6$ and the 
${\bf mf}$ for G-dwarf primaries is reduced, after ring 
dissolution, by the interactions which occur between binary 
systems formed in different rings. This is essentially what 
Kroupa (1995a) calls {\it stimulated evolution}.


\subsection{Eccentricity}
The eccentricity distribution of G-dwarfs in our model is
consistent with the observed eccentricity distribution  for 
long-period systems (DM91; Fig.~\ref{FIG:eANDq2}a). We expect that 
the eccentricities of short-period systems are modified by tidal 
forces between the two components and/or mass equalization by 
accretion (e.g. Whitworth et al., 1995; Bate, 2000), 
neither of which processes is modelled in this work. Therefore 
we only compare our model to the observed eccentricity distribution 
for long-period systems. The observations are fitted best by
${\cal N} = 4\;{\rm or}\;5$ and $\sigma_{\log M} = 0.4\;{\rm or}\;0.6$ 
(see Fig.~\ref{FIG:eANDq2}a).

For all cases, there is an excess of binaries with very high 
eccentricity ($e>0.9$) compared to the observations. The 
simulations of Delgado et al. (2003) show that these systems 
would migrate to lower eccentricities if proper account were 
taken of gas dynamical processes at periastron during the 
protostellar stage (tidal interactions, mass exchange, etc.). 
High-eccentricity systems may also be more susceptible to 
tidal disruption, as noted by Kroupa (1995b).


\subsection{Mass-ratio}
\begin{figure*}
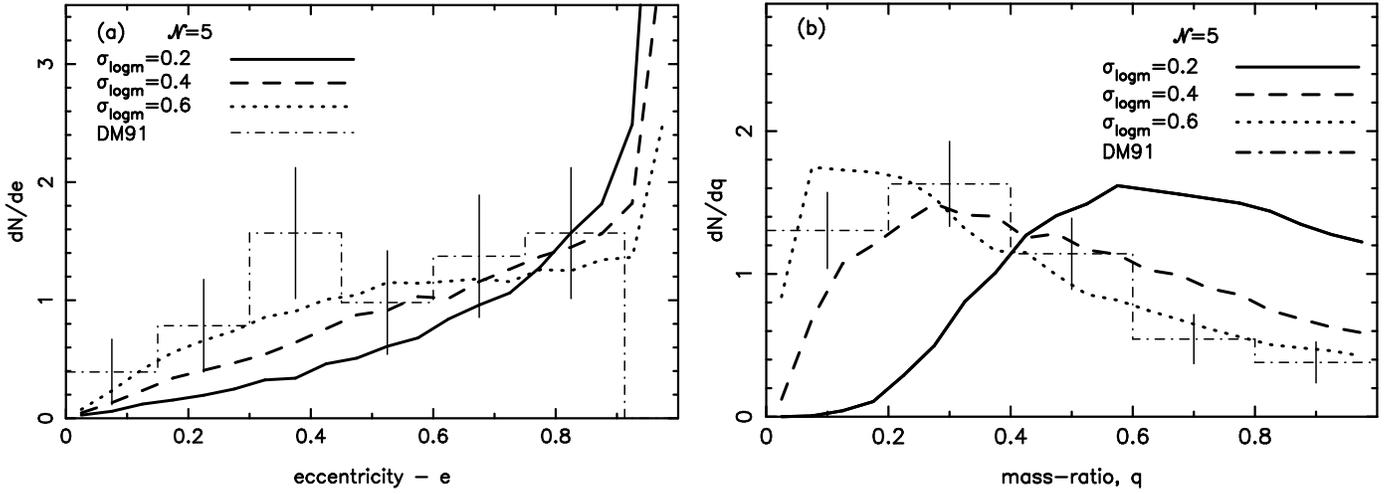
 
\centerline{\psfig{figure=2428Gecc.ps,height=6.5cm,width=9.0cm,angle=270}
\,\,\,\,\psfig{figure=2428Gq.ps,height=6.5cm,width=9.0cm,angle=270}}
\caption{The distributions of (a) eccentricity and (b) 
mass-ratio for the long-period G-dwarf binaries observed by DM91 
(dot-dashed histogram, are compared with our model results. The 
model parameters used are ${\cal N} = 5$ and $\sigma_{\log M} = 0.2\;
({\rm solid\;line}),\;0.4\;({\rm dashed\;line})\;\&\;0.6\;({\rm 
dotted\;line})$.}
\label{FIG:eANDq2}
\end{figure*}

The mass-ratio distribution from our model, with ${\cal N} = 4\;
{\rm or}\;5$ and $\sigma_{\log M} = 0.4\;{\rm or}\;0.6$, is very 
similar to the mass-ratio distribution for G-dwarfs observed by 
DM91 (as illustrated on Fig.~\ref{FIG:eANDq2}b), but somewhat 
different from the mass ratio distribution observed in Taurus by 
Woitas et al. (2001). This difference may be attributable to 
Taurus having an unusual IMF, and in particular a paucity of brown 
dwarfs (Brice\~no et al. 2002; Luhman et al. 2003). However, the 
evidence for an unusual IMF in Taurus has recently been called into 
question (Luhman 2004; Kroupa et al. 2003), and it may be necessary 
to seek another explanation for this difference.


\section{Convolving with the distribution of core rotation rates} 
\label{S:CONVBETA}

\subsection{The binary separation distribution}

In order to obtain the distribution of separations, we must convolve 
the dimensionless 
results with both the distribution of core masses $M_{_{\rm CORE}}$, 
{\it and} the distribution of core $\beta$-values. $M_{_{\rm CORE}}$ 
gives the core radius $R_{_{\rm CORE}}$ through Eqn. (\ref{EQN:Larson}),
and $\beta$ gives the ring radius $R_{_{\rm RING}}$ through Eqn. 
(\ref{EQN:RRING}). Knowing $M_{_{\rm CORE}}$ and $R_{_{\rm RING}}$, 
the dimensionless simulations can be scaled to give the distributions 
of $a$ and $P$, as a function of primary mass $M_{_1}$. As explained 
in Section \ref{SS:COREBETA} we consider two different $\beta$ 
distributions.

Using Case A ($\overline{\log \beta} = -2.0$, $\sigma_{\log \beta} = 
1.2$), our model gives the distribution of semi-major axes illustrated 
in Figure \ref{FIG:SMA}a (dotted line), where it is compared with the 
observations of pre-Main Sequence binaries having primary mass in the 
range $0.5\,{\rm M}_\odot \leq M_{_1} < 5.0\,{\rm M}_\odot$ collated 
by Patience et al. (2002; dashed line and histogram). We see that there 
is close corresondence between these two distributions. The implication 
is that, if cores have a distribution of rotation rates similar 
to the one adopted in Case A, then the dynamical dissolution of 
small rings of protostars is able to reproduce the distribution 
of binary periods observed for pre-Main Sequence stars like 
those in Taurus and Ophiuchus.

Using Case B ($\overline{\log \beta} = -2.2$, $\sigma_{\log \beta} 
= 1.7$), our model gives the distribution of semi-major axes illustrated 
in Figure \ref{FIG:SMA}b (dotted line), where it is compared 
with the observations of G Dwarf binaries in the field by DM91 
(dotted line and histogram). We see that there is a very close 
correspondence between the two distributions. The implication 
is that, if cores have a distribution of rotation rates similar 
to the one adopted in Case B, then the dynamical dissolution of 
small rings of protostars is able to reproduce the distribution 
of binary periods observed for G-Dwarf binaries in the field by 
DM91. The problem is that if we persist with the parameters 
${\cal N} = 5$ and $\sigma_{\log M} = 0.6$, the resulting 
multiplicity fraction for G-dwarf binaries is much higher than 
observed, as noted in Section \ref{SS:CMMULT}.

However, if most of the stars in the field are formed in populous 
clusters, we must allow that their binary statistics continue to 
evolve dynamically after the dissolution of the rings in which they 
are formed. The cause of further dynamical evolution is that, 
following dissolution of the individual rings, the stars 
and binary systems formed in one ring interact with the stars and 
binary systems formed in other neighbouring rings. In other words, 
the individual rings are just subclusters within the larger cluster, 
and interactions on the scale of the cluster also influence the 
final binary statistics. Kroupa (1995a) has shown that 
interactions between binaries in clusters can widen the distribution 
of binary separations somewhat, and reduce the multiplicity from near 
unity to that observed in the field.


\subsection{Dissolution timescales} \label{SS:DISSOLVE}
The dimensionless ring dissolution times, $T_{_{\rm DISS}}/
T_{_{RING}}$, derived in Section \ref{SS:DIMDISS}, can be converted 
into physical times by combining Eqns. \ref{EQN:Larson}
and \ref{EQN:RRING} with $T_{_{RING}}$ (Section \ref{SS:DIMDISS})
to obtain
\begin{equation} \label{EQN:DECAY}
\left(\frac{T_{_{\rm DISS}}}{Myr} \right) \;=\;
\left\{ \;\begin{array}{ll}
2.8 \left( \frac{T_{_{\rm DISS}}}{T_{_{\rm RING}}} \right)\,
\left( \frac{M_{_{\rm CORE}}}{{\rm M}_\odot} \right)\;\beta^{3/2}
,\hspace{0.5cm} & M_{_{\rm CORE}} \leq   {\rm M}_\odot \,; \\
 & \\
2.8 \left( \frac{T_{_{\rm DISS}}}{T_{_{\rm RING}}} \right)\,
\left( \frac{M_{_{\rm CORE}}}{{\rm M}_\odot} \right)^{1/4}{\beta}^{3/2} 
, & M_{_{\rm CORE}} \geq  {\rm M}_\odot \,.
\end{array} \right.
\end{equation}
Hence by convolving the distribution of $T_{_{\rm DISS}}/
T_{_{RING}}$ (Fig.~\ref{FIG:DECAYTIME}) with the core mass spectrum 
(Eqn.~\ref{EQN:CMS}) and the $\beta$-distribution (Eqn.~\ref{EQN:BETAFIT}), 
we obtain the overall distribution of dissolution times for the ring-clusters 
invoked in our model (Figure \ref{FIG:DECAYTIME}b). The distribution peaks 
at around $\sim 0.2\,{\rm Myr}$, the majority of the ring- clusters has 
dissolved by $1\,{\rm Myr}$, and only a handful ($< 3 \%$) remains after 
$100\,{\rm Myr}$.


\section{Conclusions} \label{S:CONCLUSIONS}

We have developed a model of binary star formation in which 
cores collapse and bounce to produce rings, and the rings 
then fragment into protostars. The dynamical evolution of 
the protostars is followed using Aarseth's NBODY3 code, 
for many different realizations, and the properties of the 
resulting multiple systems are recorded.

We adopt (i) the observed distribution of core masses, 
(ii) the observed scaling relation between core mass and core 
radius, and (iii) two possible log-normal distribution of core 
rotation (Case A and Case B), which are consistent with 
the limited observations of core rotation, but attempt to take 
account of the selection effects which make the rotation of 
some cores unmeasurable (see Fig.~\ref{FIG:BETA}). Case A 
($\overline{\log \beta} = -2.0$, $\sigma_{\log \beta} = 1.2$), 
which is our prefered option, assumes that the tendency to 
detect higher rotation rates is small, i.e. most of the 
undetected rotation rates are not much smaller than the 
detected ones. In contrast, Case B ($\overline{\log \beta} 
= -2.2$, $\sigma_{\log \beta} = 1.7$) assumes that the tendency 
to detect higher rotation rates is large, i.e. the 
undetected rotation rates are usually much smaller than the 
detected ones. We stress that this dichotomy of model rotation 
rates is unavoidable, because of the incompleteness of the 
observed rotation rates, {\it but} it only affects the 
distribution of binary separations and the time-scale on 
which ring dissolution takes place.

{\it Irrespective of the distribution of core 
rotation rates,} our model reproduces the observed IMF 
(Fig.~\ref{FIG:IMF}), and the distributions of eccentricity 
and mass ratio in {\it long-period} binary systems 
(Fig.~\ref{FIG:eANDq2}), provided only that (a) each ring 
spawns ${\cal N} \sim 5$ protostars, and (b) the protostars 
have a log-normal mass-distribution with standard deviation 
$\sigma_{\log M} \sim 0.6$. Thus ${\cal N} \sim 5$ and 
$\sigma_{\log M} \sim 0.6$ are our prefered choices for 
these free parameters. The distributions of eccentricity 
and mass ratio for {\it short-period} binary systems are not 
reproduced by our model, but we presume that this is because 
our model does not include tidal circularization or mass 
equalization by accretion.

The model also reproduces the observed variation of 
multiplicity with primary mass for mature field stars 
(Fig.~\ref{FIG:MULT}), but only if $\sigma_{\log M} \sim 0.2$. 
With our prefered values, ${\cal N} \sim 5$ and 
$\sigma_{\log M} \sim 0.6$, the model 
produces multiplicities which are higher than those observed 
for mature field stars, but are very similar to those observed 
for pre-Main Sequence stars, in the mass range 
$0.5\;{\rm to}\;5\,{\rm M}_\odot$ (this is the only mass range 
for which reliable pre-Main Sequence multiplicities are available). 
We conclude that the multiplicities resulting from ring 
dissolution represent the observed pre-Main Sequence population 
well, and that the pre-Main Sequence population then evolves 
into the observed Main Sequence population 
through interactions between stars and binary systems from different 
rings in the same cluster.

In a somewhat different context, Kroupa 
(1995a) has shown that such interactions can reduce the overall 
multiplicity from $\sim 1.0$ to $\sim 0.6$. Our model only requires 
such interactions to destroy $\sim 25\%$ of pre-Main Sequence binaries, 
but a concern with the model is then that binaries with low mass-ratio 
will be destroyed preferentially (although not exclussively). This will skew 
the distribution of mass-ratios somewhat, and may thereby degrade the agreement 
with observations of mature G-dwarfs in the field. We are currently 
conducting numerical experiments to explore the effect of interactions 
between binary systems from different rings, in the cluster environment.

In order to predict the distribution of semi-major axes, 
we have to consider the distribution of core rotation rates.
If we adopt Case A, our results match closely the distribution 
of semi-major axes obtained by Patience et al. (2002) 
for pre-Main Sequence stars (Fig. \ref{FIG:SMA}a), 
and we conclude that further dynamical evolution takes place 
-- presumably involving interactions with protostars from 
other neighbouring rings -- to convert this distribution 
into the one observed in the field.

Conversely, if we adopt Case B, our results match closely the 
peak and width of the separation distribution obtained by DM91 for 
field G Dwarfs (Fig. \ref{FIG:SMA}a). However, our results do not 
match the observed multiplicity frequency for field G-dwarfs, in 
the sense that our predicted ${\bf mf}$-value is too high. It is 
unlikely that further dynamical evolution, following 
ring dissolution, can rectify this, since to preserve the peak 
and width of the separation distribution would require dynamical 
processes which destroy, with equal efficiency, binary systems 
having widely different separations. Therefore Case A is our 
prefered option. 

Most rings have dissolved after $1\,{\rm Myr}$ and fewer than 
$3\,\%$ remain after $100\,{\rm Myr}$.


\begin{acknowledgements}
We would like to thank Simon Goodwin for many helpful discussions 
during this work, Gary Fuller for useful comments on the rotation 
of cores, and Sverre Aarseth for his NBODY3 code and patient help 
with learning to use it. We also thank the referee for comments 
which helped to improve the original version of the paper. DAH 
gratefully acknowledges the support 
of a PPARC postgraduate studentship, and APW acknowledges the 
support of a European Commission Research Training Network awarded 
under the Fifth Framework (Ref. HPRN-CT-2000-00155).
\end{acknowledgements}


\appendix\section{Skewing the log-normal distribution of stellar masses}

Stellar masses, $M_n\;\,(n = 1,{\cal N})\,$ are initially  picked 
randomly from a log-normal distribution which is symmetric in 
${\rm log}M$ (Eqn. \ref{MS}). These masses are then re-scaled by a 
factor $g$, $\;M_n \rightarrow \mu_n = g M_n\,$, so that the total 
mass is unity (Eqn. \ref{EQN:NORM}). After this re-scaling, the 
minimum possible mass is 0, corresponding to ${\rm log}M = - \infty$, 
and the maximum possible mass is 1, corresponding to ${\rm log}M = 0$. 
It follows that the distribution can now only be symmetric in 
${\rm log}M$ if its median lies at ${\rm log}M = - \infty$, but this 
is clearly non-sensical. In fact the distribution is skewed, in the 
sense of having a tail to low values of $\mu$, and the median 
is of order $\,-\,{\rm log}({\cal N})\,$. 

The asymmetry arises because we are invoking a finite -- indeed small 
-- number of stars. For simplicity consider the case ${\cal N} = 2\,$, 
and assume that the two stars do not have equal mass. If the more 
massive star is initially (pre re-scaling) exceptionally massive, it 
has to be reduced to achieve $\sum\{ \mu_n \} = 1\,$, and this 
inevitably decreases the mass of the lower mass star, even if it 
was already quite small. Conversely, if the less massive star is 
initially of exceptionally low-mass, this has little or no influence 
on the re-scaling, which is mainly influenced by the more massive star. 
It is this asymmetry between the effects of exceptionally high-mass 
and exceptionally low-mass stars on the re-scaling that causes the 
re-scaled distribution to be skewed, particularly for small 
${\cal N}$. The effect disappears as ${\cal N} \rightarrow \infty$.


\end{document}